\documentclass[prd,superscriptaddress,unsortedaddress,twocolumn,showpacs,preprintnumbers,amsmath,amssymb]{revtex4}

\usepackage[dvips,dvipdfmx]{graphicx}
\usepackage{amsmath,amssymb,times}

\def\lessa{\mathrel{\mathpalette\fun <}}
\def\greata{\mathrel{\mathpalette\fun >}}
\def\fun#1#2{\lower3.6pt\vbox{\baselineskip0pt\lineskip.9pt
\ialign{$\mathsurround=0pt#1\hfil##\hfil$\crcr#2\crcr\sim\crcr}}}

\newcommand{\beq}{\begin{equation}}
\newcommand{\eeq}{\end{equation}}
\newcommand{\bea}{\begin{eqnarray}}
\newcommand{\eea}{\end{eqnarray}}

\renewcommand{\a}{\alpha}

\DeclareSymbolFont{boldletters}{OML}{cmm} {b}{it}
\DeclareSymbolFontAlphabet{\mathbit}{boldletters}
\DeclareMathSymbol{\alpha}{\mathalpha}{letters}{"0B}
\DeclareMathSymbol{\beta}{\mathalpha}{letters}{"0C}
\DeclareMathSymbol{\gamma}{\mathalpha}{letters}{"0D}
\DeclareMathSymbol{\delta}{\mathalpha}{letters}{"0E}
\DeclareMathSymbol{\epsilon}{\mathalpha}{letters}{"0F}
\DeclareMathSymbol{\zeta}{\mathalpha}{letters}{"10}
\DeclareMathSymbol{\eta}{\mathalpha}{letters}{"11}
\DeclareMathSymbol{\theta}{\mathalpha}{letters}{"12}
\DeclareMathSymbol{\iota}{\mathalpha}{letters}{"13}
\DeclareMathSymbol{\kappa}{\mathalpha}{letters}{"14}
\DeclareMathSymbol{\lambda}{\mathalpha}{letters}{"15}
\DeclareMathSymbol{\mu}{\mathalpha}{letters}{"16}
\DeclareMathSymbol{\nu}{\mathalpha}{letters}{"17}
\DeclareMathSymbol{\xi}{\mathalpha}{letters}{"18}
\DeclareMathSymbol{\pi}{\mathalpha}{letters}{"19}
\DeclareMathSymbol{\rho}{\mathalpha}{letters}{"1A}
\DeclareMathSymbol{\sigma}{\mathalpha}{letters}{"1B}
\DeclareMathSymbol{\tau}{\mathalpha}{letters}{"1C}
\DeclareMathSymbol{\upsilon}{\mathalpha}{letters}{"1D}
\DeclareMathSymbol{\phi}{\mathalpha}{letters}{"1E}
\DeclareMathSymbol{\chi}{\mathalpha}{letters}{"1F}
\DeclareMathSymbol{\psi}{\mathalpha}{letters}{"20}
\DeclareMathSymbol{\omega}{\mathalpha}{letters}{"21}
\DeclareMathSymbol{\varepsilon}{\mathalpha}{letters}{"22}
\DeclareMathSymbol{\vartheta}{\mathalpha}{letters}{"23}
\DeclareMathSymbol{\varpi}{\mathalpha}{letters}{"24}
\DeclareMathSymbol{\varrho}{\mathalpha}{letters}{"25}
\DeclareMathSymbol{\varsigma}{\mathalpha}{letters}{"26}
\DeclareMathSymbol{\varphi}{\mathalpha}{letters}{"27}
\DeclareMathSymbol{\Gamma}{\mathalpha}{letters}{"00}
\DeclareMathSymbol{\Delta}{\mathalpha}{letters}{"01}
\DeclareMathSymbol{\Theta}{\mathalpha}{letters}{"02}
\DeclareMathSymbol{\Lambda}{\mathalpha}{letters}{"03}
\DeclareMathSymbol{\Xi}{\mathalpha}{letters}{"04}
\DeclareMathSymbol{\Pi}{\mathalpha}{letters}{"05}
\DeclareMathSymbol{\Sigma}{\mathalpha}{letters}{"06}
\DeclareMathSymbol{\Upsilon}{\mathalpha}{letters}{"07}
\DeclareMathSymbol{\Phi}{\mathalpha}{letters}{"08}
\DeclareMathSymbol{\Psi}{\mathalpha}{letters}{"09}
\DeclareMathSymbol{\Omega}{\mathalpha}{letters}{"0A}

\begin{document}
\title{Determination of the strength of the vector-type four-quark 
interaction \\
in the entanglement Polyakov-loop extended Nambu--Jona-Lasinio model} 

\author{Junpei Sugano}
\email[]{sugano@phys.kyushu-u.ac.jp}
\affiliation{Department of Physics, Graduate School of Sciences, Kyushu University,
             Fukuoka 812-8581, Japan}             

\author{Junichi Takahashi}
\email[]{takahashi@phys.kyushu-u.ac.jp}
\affiliation{Department of Physics, Graduate School of Sciences, Kyushu University,
             Fukuoka 812-8581, Japan}

\author{Masahiro Ishii}
\email[]{ishii@phys.kyushu-u.ac.jp}
\affiliation{Department of Physics, Graduate School of Sciences, Kyushu University,
             Fukuoka 812-8581, Japan}

\author{Hiroaki Kouno}
\email[]{kounoh@cc.saga-u.ac.jp}
\affiliation{Department of Physics, Saga University,
             Saga 840-8502, Japan}  

\author{Masanobu Yahiro}
\email[]{yahiro@phys.kyushu-u.ac.jp}
\affiliation{Department of Physics, Graduate School of Sciences, Kyushu University,
             Fukuoka 812-8581, Japan}

\date{\today}

\begin{abstract}
We determine the strength $G_{\rm v}$ of the vector-type four-quark 
interaction in the entanglement Polyakov-loop extended Nambu--Jona-Lasinio (EPNJL) 
model from the results of recent lattice QCD simulations with 
two-flavor Wilson fermion. 
The quark number density is normalized by the Stefan-Boltzmann limit. 
The strength determined from the normalized quark number density 
at a baryon chemical potential $\mu$ and temperature $T$ (which is higher 
than the pseudocritical temperature $T_c$ of the deconfinement transition) 
is $G_{\rm v}=0.33 G_{\rm s}$, where $G_{\rm s}$ 
is the strength of the scalar-type four-quark interaction. 
We explore the hadron-quark phase transition in the $\mu$--$T$ plane by 
using the two-phase model in which the quantum hadrodynamics model is 
used for the hadron phase and the EPNJL model is used for the quark phase. 
When $G_{\rm v}=0.33 G_{\rm s}$, the critical baryon chemical potential 
of the transition at zero $T$ is $\mu_c \sim 1.6$ GeV, which accounts 
for two-solar-mass measurements of neutron stars 
in the framework of the hadron-quark hybrid star model.    
\end{abstract}

\maketitle

{\it Introduction.} 
Determining the QCD phase diagram is an important subject 
in particle and nuclear physics, as well as in cosmology and astrophysics. 
However, using lattice QCD (LQCD) as a first-principle calculation presents 
a sign problem at finite baryon chemical potential $\mu$. 
Several methods were proposed to resolve this problem, 
such as the reweighting method~\cite{Fodor}, the Taylor expansion method~\cite{Allton,Ejiri_density}, and the analytic continuation from imaginary $\mu$ 
to real $\mu$~\cite{FP,D'Elia,D'Elia3,FP2010,Nagata,Takahashi,Chen}. 
The results are reliable for small $\mu$; say, $\mu/T \lessa 3$, where $T$ is 
the temperature.
Very recently, remarkable progress toward larger $\mu/T$ 
has been made 
with the complex Langevin method~\cite{Aarts_CLE_1,Aarts_CLE_2,Sexty} 
and the  Lefschetz thimble theory~\cite{Aurora_thimbles,Fujii_thimbles}.  
However, the results are still far from the physics at $\mu/T = \infty$, 
as occurs in nuclear matter and neutron stars. 

In another important approach, one can consider effective models such as 
the Polyakov-loop extended Nambu--Jona-Lasinio (PNJL) model 
~\cite{Meisinger,Dumitru,Fukushima1,Ghos,Megias,Ratti1,Rossner}.  
The PNJL model can treat the confinement mechanism approximately and the chiral symmetry breaking. 
As for zero $T$, 
the chiral and deconfinement transitions coincide with each other 
in LQCD simulations, but not in the PNJL model, 
when the model parameters are set to 
the realistic transition temperature $T_c$~\cite{Ratti1}. 
This problem was solved by introducing the Polyakov-loop dependent 
four-quark interaction to the PNJL model~\cite{Sakai_EPNJL,Sasaki_EPNJL}. 
This model is called the entanglement-PNJL (EPNJL) model. 
The EPNJL model also accounts for the phase structure 
calculated with LQCD at imaginary $\mu$~\cite{D'Elia3,FP2010} and 
real isospin chemical potential~\cite{Kogut2}. 
Very recently, Ishii {\it et al.}~\cite{Ishii} showed that 
the EPNJL model reproduces the meson screening masses on 
temperature, as calculated with LQCD~\cite{Cheng_sc_mass}. 

Since baryons are not treated in NJL-type models, 
a plausible approach is to take 
the two-phase model in which two different models are used between the hadron and quark phases to analyze the hadron-quark transition.  
In NJL-type models for the quark phase, 
the stiffness of the equation of state (EoS) is sensitive 
to the strength $G_{\rm v}$ of the vector-type four-quark 
interaction~\cite{Kashiwa1,Sakai_vector}. 
Reference~\cite{Sasaki_NS} shows that 
the condition $G_{\rm v}\ge 0.03G_{\rm s}$ is necessary 
for neutron stars (NSs)
to have masses larger than two solar masses ($2 M_\odot $), where 
$G_{\rm s}$ is the strength of the scalar-type four-quark interaction. 
If $G_{\rm v} < 0.03G_{\rm s}$, the EoS for the hadron phase is softened by the hadron-quark transition before exceeding two solar masses. 
The value of $G_{\rm v}$ is thus quite important to explain the observed 
$2 M_\odot $ NSs~\cite{Demorest,Antoniadis}. 

Sakai {\it et al.}~\cite{Sakai_para} estimated the strength of $G_{\rm v}$ 
from two-flavor LQCD results \cite{FP,Chen} 
for the deconfinement transition line 
at the imaginary chemical potential by using the PNJL model 
with the vector-type four-quark and scalar-type eight-quark interactions 
in addition to the scalar-type four-quark 
interaction. 
Applying the mean-field approximation reduces the scalar-type four- 
and eight-quark interactions to a scalar-type four-quark interaction 
with an effective strength $G_{\rm s}^*$. They suggested 
that $G_{\rm v}/G_{\rm s}^*\approx 0.8$. 
A similar analysis based on the nonlocal PNJL model suggests that 
$G_{\rm v}/G_{\rm s} \approx 0.4$~\cite{Kashiwa_nonlocal}.   
Recently, using the PNJL-like model, 
Steinheimer and Schramm~\cite{Steinheimer} estimated 
the strength of $G_{\rm v}$ from three-flavor LQCD results~\cite{Borsanyi_2} 
for the quark number susceptibility, and concluded 
that $G_{\rm v}$ is nearly zero. Meanwhile, Lourenco {\it et al.}
estimated the strength of $G_{\rm v}$ by comparing 
the PNJL model with the two-phase model 
for the hadron-quark phase transition and suggested that 
$1.52 \lesssim G_{\rm v}/G_{\rm s}\lesssim 3.2$~\cite{Lourenco}. 
The strength of $G_{\rm v}$ is thus undetermined. 

Because previous analyses are mainly based on LQCD simulations 
with the Kogut-Susskind fermion, 
we determine in this brief report the strength of $G_{\rm v}$ 
in the EPNJL model from the results 
of recent two-flavor LQCD simulations~\cite{Ejiri_density} 
with the Wilson fermion at $T > T_c$ and small $\mu/T$.  
The quark number density $n_q$ is sensitive 
to the strength of $G_{\rm v}$, but is $\mu$ odd and thus tiny 
for small $\mu$. 
It is then convenient to consider the quark number density normalized 
by the Stefan-Boltzmann (SB) limit, $n_q/n_{\rm SB}$. 
The normalized quark number density is $\mu$ even 
and thus finite even in the limit of $\mu=0$. 
It hardly depends on $\mu$ in the region $\mu/T \lessa 1$ 
where LQCD data~\cite{Ejiri_density} are available. 
The ratio $n_q/n_{\rm SB}$ is considered to be more reliable 
in the vicinity of $\mu =0$, because the results there are obtained with 
the Taylor-expansion method. 
Therefore, we consider the ratio $n_q/n_{\rm SB}$ in the limit 
$\mu \rightarrow 0$ 
to estimate the strength of $G_{\rm v}$. 
We show herein that the strength of $G_{\rm v}$ 
thus determined is $G_{\rm v}=0.33 G_{\rm s}$ and is not so small.

We also draw the hadron-quark transition line in the $\mu$-$T$ plane by 
using the two-phase model composed of the EPNJL model with the vector-type 
interaction for the quark phase and the quantum hadrodynamics (QHD) model 
for the hadron phase. 
We evaluate the critical baryon chemical potential $\mu_c$ of the transition 
at $T=0$ and discuss whether 
the result for $\mu_c$ is consistent with observations of $2 M_\odot $ NSs.

{\it EPNJL model for quark phase.} 
We add the vector-type four-quark interaction to 
the isospin-symmetric two-flavor EPNJL model~\cite{Sakai_EPNJL,Sasaki_EPNJL}. 
The Lagrangian density is 
\begin{align}
 {\cal L}_{\rm EPNJL}  
=& {\bar q}(i \gamma^\mu D_\mu -m_0)q  + \tilde{G}_{\rm s}(\Phi)[({\bar q}q )^2 
  +({\bar q }i\gamma_5 {\vec \tau}q )^2]
\nonumber\\
 & -\tilde{G}_{\rm v}(\Phi )({\bar q \gamma_\mu q})^2
 -{\cal U}(\Phi [A],{\Phi}^* [A],T) ,
\label{L}
\end{align}
where $q$ is the quark field, $m_0$ is the current quark mass,  
and ${\vec \tau}$ is the isospin matrix. 
As a characteristic of the EPNJL model, the coupling constants 
$\tilde{G}_{\rm s}(\Phi)$ and $\tilde{G}_V(\Phi)$ of the 
scalar- and vector-type four-quark interactions depend 
on the Polyakov-loop $\Phi$,
\begin{eqnarray}
\tilde{G}_{\rm s}(\Phi)&=&G_{\rm s}\left[1-\alpha_1\Phi{\Phi}^* -\alpha_2\left(\Phi^3 + {\Phi^*}^{3}\right)\right],
\label{EPNJL_GS}
\\
\tilde{G}_{\rm v}(\Phi )&=&
G_{\rm v} \left[1-\alpha_1\Phi{\Phi}^* -\alpha_2\left(\Phi^3 + {\Phi^*}^{3}\right)\right] ,
\nonumber\\
\label{EPNJL_GV} 
\end{eqnarray}
where 
$D^\mu=\partial^\mu+iA^\mu$ for $A^\mu=g\delta^{\mu}_{0}(A^0)_a{\lambda_a/2}
=-ig\delta^{\mu}_{0}(A_{4})_a{\lambda_a/2}$, $A^\mu_a$ is the gauge field, 
$\lambda_a$ is the Gell-Mann matrix, and $g$ is the gauge coupling. 
Eventually, the NJL sector has five parameters 
$(m_0, G_{\rm s},G_{\rm v}, \alpha_1, \alpha_2)$. 
In the present parametrization, the ratio 
$\tilde{G}_{\rm v}(\Phi)/\tilde{G}_{\rm s}(\Phi)$ is independent of $\Phi$, 
and $\tilde{G}_{\rm s}(\Phi )=G_{\rm s}$ 
and $\tilde{G}_{\rm v}(\Phi )=G_{\rm v}$ at $T=0$ where $\Phi=0$. 
When $\alpha_1=\alpha_2=0$, 
the EPNJL model reduces to the PNJL model.

In the EPNJL model, only the time component of $A_\mu$ is treated as a homogeneous and static background field. This parameter is governed by the Polyakov-loop potential~~$\mathcal{U}$. 
The Polyakov-loop $\Phi$ and its conjugate ${\Phi}^*$ are then obtained in the Polyakov gauge as 
\begin{align}
\Phi &= {1\over{3}}{\rm tr}_{\rm c}(L),
~~~~~{\Phi}^* ={1\over{3}}{\rm tr}_{\rm c}({L^\dagger}) ,
\label{Polyakov}
\end{align}
where $L= \exp[i A_4/T]=\exp[i~{\rm Diag}(A_4^{11},A_4^{22},A_4^{33})/T]$ 
for the classical variables $A_4^{ii}$ 
satisfying $A_4^{11}+A_4^{22}+A_4^{33}=0$. 
We use the logarithm-type Polyakov-loop potential $\mathcal{U}$ of Ref.~\cite{Rossner}. 
The parameter set in $\mathcal{U}$ is fit to reproduce LQCD data at finite $T$ in the pure gauge limit. 
The potential $\mathcal{U}$ yields 
the first-order deconfinement phase transition 
at $T=T_0$. In the pure gauge limit, 
LQCD reveals a phase transition at $T=270$~MeV.  
Thus, the parameter $T_0$ is often set to $270$~MeV; however, with this value of $T_0$, the EPNJL model yields a larger $T_\mathrm{c}$ for the deconfinement transition than the full-LQCD prediction $T_{\rm c} = 173\pm 8$~MeV~\cite{Borsanyi,Soeldner,Kanaya}. 
We thus rescale $T_0$. The EPNJL model with $T_0=190$~MeV and $\alpha_1=\alpha_2=0.2$ reproduces well the full LQCD results for the deconfinement and chiral 
transition lines at zero and imaginary $\mu$~\cite{Sakai_EPNJL}. 
As mentioned above, the parameter $\alpha_3 \equiv G_{\rm v}/G_{\rm s}$ 
is determined from the full LQCD results for $n_q/n_{\rm SB}$ 
in the limit $\mu \rightarrow 0$.

For the NJL sector, we take the same parameter set as 
in Ref.~\cite{Sakai_EPNJL} except for the current quark mass $m_0$. 
The LQCD simulations of Ref.~\cite{Ejiri_density} were done 
on a $4 \times 16^3$ lattice with two-flavor clover-improved Wilson quark action along the line of 
constant physics of $m_{\pi}/m_{\rho}=0.65$ and 0.8 for 
$\pi$- and $\rho$-meson masses $m_{\pi}$ and $m_{\rho}$, respectively. 
The corresponding vacuum values of $m_{\pi}$ are 500 MeV and 616 MeV, and 
the parameters refit to these values are $m_0=72$~MeV and 130~MeV.

The mean field approximation to Eq. \eqref{L} leads to 
the thermodynamic potential (per unit volume) of 
\begin{align}
&\Omega_{\rm EPNJL} \nonumber\\ 
&= U_{\rm M}+{\cal U}-2 N_{\rm f} \int \frac{d^3 p}{(2\pi)^3}
   \Bigl[ 3 E \notag \\
&+ \frac{1}{\beta}
           \ln~ [1 + 3(\Phi+{\Phi}^* e^{-\beta (E-\tilde{\mu}_q )}) 
           e^{-\beta (E-\tilde{\mu}_q)}+ e^{-3\beta (E-\tilde{\mu}_q)}] 
\notag\\
&+ \frac{1}{\beta} 
           \ln~ [1 + 3({\Phi}^*+{\Phi e^{-\beta (E+\tilde{\mu}_q)}}) 
              e^{-\beta (E+\tilde{\mu}_q)}+ e^{-3\beta (E+\tilde{\mu}_q)}]
	      \Bigl] ,
	      \nonumber\\
\label{PNJL-Omega}
\end{align}
where $E=\sqrt{{\bf p}^2+M^2}$, $M=m_0-2\tilde{G}_{\rm s}\sigma$, 
$\tilde{\mu}_q=\mu_q -2\tilde{G}_{\rm v} n_{q}$, 
$U_{\rm M}= \tilde{G}_{\rm s}\sigma^2-\tilde{G}_{\rm v} n_q^2$, 
$N_{\rm f}$ is the number of flavors, and the quark chemical potential 
$\mu_q$ is related to the baryon chemical potential $\mu$ as $\mu=3\mu_q$.

\begin{figure}[t]
\begin{center}
\includegraphics[width=0.45\textwidth,clip]{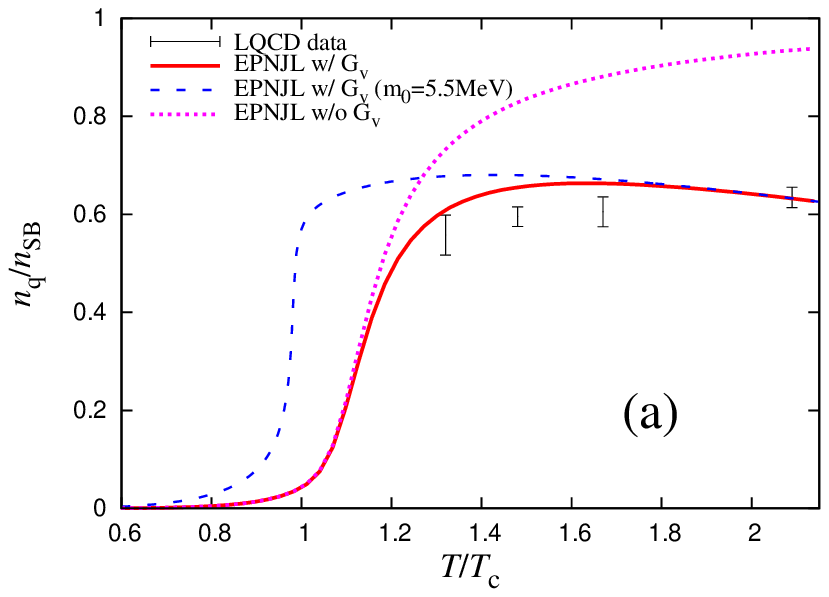}
\includegraphics[width=0.45\textwidth,clip]{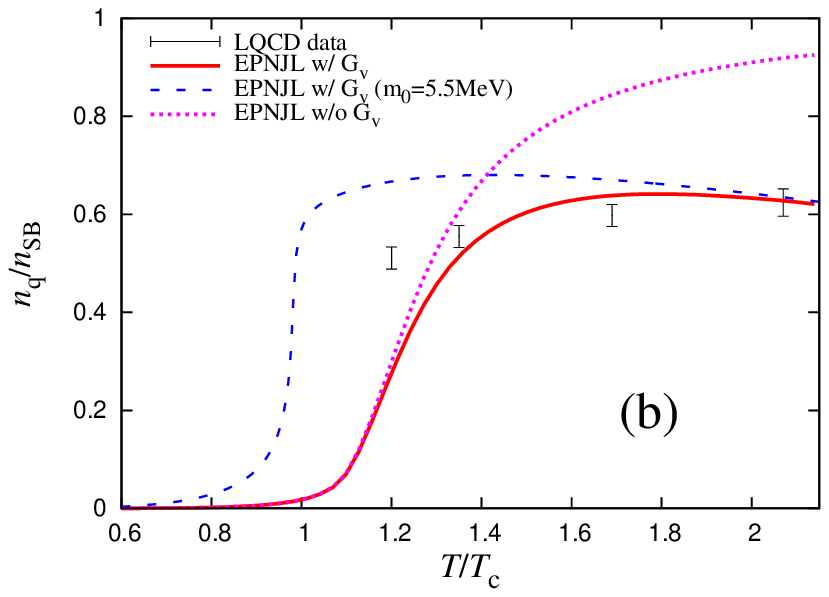}
\end{center}
\caption{The ratio $n_q/n_{\rm SB}$ 
as a function of temperature in the limit $\mu \rightarrow 0$ 
for (a) $m_0=72$ MeV and (b) $m_0=130$MeV. 
LQCD data in panels (a) and (b) are taken from the cases 
of $m_\pi /m_\rho =0.65$ and 0.8 in Ref.~\cite{Ejiri_density}, respectively. 
The dotted and solid lines denote the results of the EPNJL model 
with $G_{\rm v}=0$ and $G_{\rm v}=0.33G_{\rm s}$, respectively, 
and the dashed line represents the result of the EPNJL model 
with $G_{\rm v}=0.33G_{\rm s}$ and $m_0=5.5$ \rm{MeV}. We take the full QCD prediction as
$T_{\textrm{c}}$
}
\label{Fig-qnumber}
\end{figure}

Figures \ref{Fig-qnumber}(a) and \ref{Fig-qnumber}(b) show the $T$ dependence of $n_q/n_{\rm SB}$ in the limit $\mu_q\to 0$ for $m_0=72$~MeV and 
$m_0=130$~MeV, respectively. 
In model calculations, $n_q$ is divided by the SB limit in the 
continuum theory. 
In LQCD simulations~\cite{Ejiri_density}, $n_q$ is normalized by 
the lattice SB limit to eliminate finite-volume effects. 
The dotted and solid lines represent the EPNJL results with $G_{\rm v}=0$ and 
$G_{\rm v}=0.33G_{\rm s}$, respectively. 
In the region $1< T/T_c \lessa 1.2$, the 
$n_q/n_{\rm SB}$ depends weakly on the strength of $G_{\rm v}$. 
It is thus not easy to precisely determine the strength for $T$ near $T_c$. 
This implies that the phase-transition line 
is not a good quantity to determine the strength. 
One can see from the region $T/T_c \greata 1.2$ 
that $G_{\rm v}=0.33G_{\rm s}$ is the best value to explain the LQCD results. 
Good consistency results for both $m_0=72$ and $130$~MeV; 
therefore, the ratio $\a_3=G_{\rm v}/G_{\rm s}$ depends only weakly 
on the value of $m_0$. 

The dashed line represents the result of the EPNJL model 
with $G_{\rm v}=0.33G_{\rm s}$ in which 
$m_0$ is set to the physical value $5.5$ MeV. 
The dashed line is consistent with the solid line and with LQCD data 
for $T/T_c \greata 1.7$, where $m_0/T$ is negligibly small. 
This result means 
that the strength of $G_{\rm v}$ is clearly determined from LQCD 
data for $T/T_c \greata 1.7$, even if $m_0$ is larger than $5.5$ MeV in 
the LQCD calculations.

{\it QHD model for hadron phase.} 
We now explore the hadron-quark transition by using the value of $G_{\rm v}$ 
determined above. 
Because the EPNJL model is designed to treat 
the deconfinement transition only approximately, 
the hadron degrees of freedom are not correctly included in the model. 
We thus use the two-phase model in which the transition line is determined 
from the Gibbs criteria. 
For the hadron phase, we use the quantum QHD model of Ref. 
\cite{Lalazissis}. The Lagrangian density is 
\bea
&&{\cal L}_{\rm QHD}=\bar{\psi}(i\gamma^\mu \partial_\mu -m_N -g_\sigma \varphi -g_\omega \gamma^\mu\omega_\mu )\psi
+{1\over{2}}\partial^\mu \varphi \partial_\mu \varphi
\nonumber\\
&&\ \ \ \ \ \ \ \ \ \ \ \ \ \ \ -{1\over{4}}(\partial^\mu \omega^\nu-\partial^\nu \omega^\mu) (\partial_\mu \omega_\nu-\partial_\nu \omega_\mu)
-U_{\rm QHD},~~~~
\label{QHD}\\
&&U_{\rm QHD}={1\over{2}}m_\sigma^2\varphi^2+{1\over{3}}g_2\varphi^3+{1\over{4}}g_3\varphi^4
-{1\over{2}}m_\omega^2 \omega^\mu \omega_\mu ,~~~~
\label{QHD_sigma}
\eea
where $\psi$, $\varphi$, $\omega_\mu$, $m_N$, $m_\sigma$, and $m_\omega$ are 
nucleon (N), $\sigma$-meson and $\omega$-meson fields, and 
their masses, respectively, whereas 
$g_\sigma$, $g_\omega$, $g_2$, and $g_3$ are 
$\sigma$-N, $\omega$-N, and higher-order couplings, respectively. 
The mean field approximation to Eq. \eqref{QHD} thus yields 
the following thermodynamic potential (per unit volume): 
\begin{align}
&\Omega_{\rm QHD}
=\ \  U_{\rm QHD}(\varphi,\omega_0)-2\sum_{N=p,n}\int \frac{d^3 p}{(2\pi)^3}
   \Bigl[\notag \\
&\frac{1}{\beta}
           \ln~ [1 + e^{-\beta (E_{N}-\mu^\ast)}] 
+ \frac{1}{\beta} 
           \ln~ [1 + e^{-\beta (E_{N}+\mu^\ast)}]
	      \Bigl], 
\label{QHD-Omega}
\end{align}
for $E_N=\sqrt{{\textbf p}^2+{m_N^*}^2}$ with $m_N^*=m_N+g_\sigma \varphi$, 
$\mu^*=\mu-g_{\omega}\omega_0$. 
The meson fields were replaced by constant values in Eq. (\ref{QHD-Omega}) 
so that the spatial components of $\omega_\mu$ and all 
the kinetic terms vanished. 
Unlike in Eq. (\ref{PNJL-Omega}), the vacuum contribution term is 
not included in Eq. (\ref{QHD-Omega}), because the effects were already 
included in the physical hadron masses and couplings in the Lagrangian 
(\ref{QHD}). We use the NL3 set~\cite{Lalazissis} as 
the parameter set of the QHD model. 
For the quark phase, we use the EPNJL model with $m_0=5.5$ MeV and 
$G_{\rm v}=0.33G_{\rm s}$.

The Gibbs criteria dictate that the phase with higher pressure occurs 
between two phases. At $T=\mu=0$, the pressure $P_{\rm QHD}=-\Omega_{\rm QHD}$ 
for the hadron phase is zero by definition, whereas the pressure 
$P_{\rm EPNJL}=-\Omega_{\rm EPNL}$ for the quark phase is finite because of 
the vacuum term. 
To eliminate the ambiguity due to the vacuum term, 
we replace $P_{\rm EPNJL}$ by
\begin{eqnarray}
\tilde{P}_{\rm EPNL}(T,\mu )=P_{\rm EPNL}(T,\mu )-P_{\rm EPNL}(0,0)-B ,
\label{bag_constant}
\end{eqnarray}
which introduces the bag constant $B$ with 
$\tilde{P}_{\rm EPNJL}=-B$ at $T=\mu =0$. 
The value of $B$ is determined to reproduce the LQCD prediction of 
the pseudocritical temperature of the deconfinement transition at $\mu =0$.

\begin{figure}[t]
\begin{center}
\includegraphics[width=0.4\textwidth]{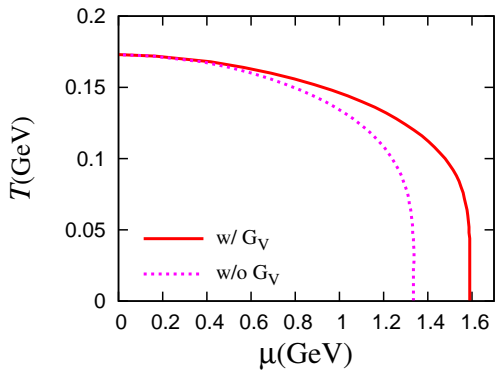}
\end{center}
\caption{Phase diagram in $\mu$-$T$ plane for hadron-quark 
phase transition.  
The solid and dotted lines show the results of the two-phase model with 
$G_{\rm v}=0.33G_{\rm s}$ and $G_{\rm v}=0$, respectively. 
In the EPNJL model, we use $m_0=5.5$MeV. We take the approximation of 
$\Phi = \Phi^{\ast}$
}
\label{phase-diagram}
\end{figure}

Figure~\ref{phase-diagram} shows the phase diagram in the $\mu$-$T$ plane 
for the hadron-quark phase transition. 
The two-phase model with $G_{\rm v}=0.33G_{\rm s}$ (solid line) shows that 
the critical baryon chemical potential of the transition at $T=0$ 
is $\mu_c \sim 1.6$ GeV. 
This value is just above the lower bound $\mu_c \sim 1.6$ GeV to account 
for the observations of $2 M_\odot $ NSs~\cite{Sasaki_NS}. 
When $G_{\rm v}=0$, the critical value at zero $T$ 
is shifted down to $\mu_c \sim 1.3$ GeV, as shown by the dotted line. 
The contribution of the vector-type four-quark interaction is 
thus quite significant.

{\it Summary.} 
We determined the strength $G_{\rm v}$ of the vector-type four-quark 
interaction in the EPNJL model by using the results of LQCD simulations 
with two-flavor clover-improved Wilson quark action at small $\mu/T$.  
The results indicate that $G_{\rm v}/G_{\rm s}\sim 0.33$ best reproduces 
LQCD data for the normalized quark number density 
$n_q/n_{\rm SB}$ for small $\mu$ and $T/T_c > 1.2$.  
The value of $G_{\rm v}$ appears to be almost independent 
of the current quark mass because the EPNJL model 
with $G_{\rm v}/G_{\rm s} = 0.33$ simultaneously accounts 
for two types of LQCD data: one with $m_{\pi}/m_{\rho}=0.65$ and the other 
with $m_{\pi}/m_{\rho}=0.8$.

The ratio $G_{\rm v}/G_{\rm s} = 0.33$ is consistent 
with the result $G_{\rm v}/G_{\rm s}=0.4$ 
obtained from the phase diagram for imaginary $\mu$ 
with the nonlocal PNJL 
model~\cite{Kashiwa_nonlocal} and is not far from the ratio 
$G_{\rm v}/G_{\rm s}=0.5$ calculated 
with a local version of the gluon exchange interaction 
model~\cite{Hatsuda_1985}.

Using $G_{\rm v}=0.33G_{\rm s}$, we explored 
the hadron-quark phase transition in the $\mu$--$T$ plane. 
The critical baryon chemical potential of the transition at $T=0$ is 
$\mu_c \sim 1.6$ GeV and is just above the lower bound $\mu_c \sim 1.6$ GeV 
to account for observations of $2 M_\odot$ NSs. 
We therefore conclude that the QCD phase diagram drawn 
with the present two-phase model is consistent with LQCD data 
at small $\mu/T$ and with observations of $2 M_\odot$ NSs at $\mu/T=\infty$.

To obtain more robust information on $G_{\rm v}$, we plan 
to analyze the $\mu$ dependence of the ratio $n_q/n_{\rm SB}$ more precisely 
for imaginary $\mu$ by using LQCD simulations with two-flavor Wilson fermion.

\noindent
\begin{acknowledgments}
We thank I.-O. Stamatescu, A. Nakamura, K. Kashiwa and T. Sasaki 
for useful discussions. 
M. Y., H. K., and J. T. are supported
by Grant-in-Aid for Scientific Research (No. 26400278, No. 26400279, and No.25-3944) from the Japan Society for the Promotion of Science (JSPS). 
\end{acknowledgments}


\end{document}